\documentclass[aps, prab, amsmath,amssymb, final,10pt,
 twoside,
twocolumn,
 nofloats,
nofootinbib,
 superscriptaddress,
showkeys,
showkeywords]{revtex4-2} 
\usepackage{amsfonts}
\usepackage{amsmath}
\usepackage{graphicx}
\usepackage{dcolumn}
\usepackage{bm}

\usepackage[utf8]{inputenc}
\usepackage{amssymb, amsthm}
\usepackage{pgfplots}

\newtheoremstyle{comment}{}{}{}{}{}{:}{ }{#1}
\theoremstyle{comment}
\usepackage{booktabs}
\usepackage{hyperref}  
\usepackage{listings}
\usepackage{textcomp}
\usepackage{diagbox}
\usepackage{comment}
\usepackage{academicons}
\usepackage{orcidlink}
\definecolor{orcidlogocol}{HTML}{A6CE39}
\usepackage[percent]{overpic}  
\begin{document}

\title{\Large 
More Than Opinions: The Role of Values in Shaping Fairness and Status in the Ultimatum Game within Structured Societies
}
\author{\orcidlink{0000-0002-0062-3377}Hana Krakovsk\'{a}}
\email[Corresponding author:]{hana.krakovska@savba.sk}
\affiliation{Institute of the Science of Complex Systems, Center for Medical Data Science, Medical University of Vienna, Spitalgasse 23, Vienna, 1090, Austria}
\affiliation{Complexity Science Hub, Metternichgasse~8, Vienna, 1030, Austria}
\affiliation{Institute of Measurement Science, Slovak Academy of Sciences, D\'{u}bravsk\'{a} cesta 9, Bratislava, Slovakia}
\author{\orcidlink{0000-0002-4045-9532}Rudolf Hanel} 
\affiliation{Institute of the Science of Complex Systems, Center for Medical Data Science, Medical University of Vienna, Spitalgasse 23, Vienna, 1090, Austria}
\affiliation{Complexity Science Hub, Metternichgasse~8, Vienna, 1030, Austria}
\date{Version \today}
    \begin{abstract}
        Asymmetric evolutionary 
        games, such
        as the Ultimatum Game, provide keys to understanding the emergence of fairness in social species. 
         Building on this framework, we explore the evolution of social value systems and the operational role that social status plays in hierarchically organised societies.
      Within the asymmetric Ultimatum Game paradigm, where "proposers" suggest terms for resource distribution, and "responders" accept or reject these terms, we examine the assignment of roles between players under a subjective social order.
        This order is grounded in an emergent status hierarchy based on observable player attributes (such as age and wealth). The underlying rules for constructing such a hierarchy stabilise over time by inheritance and family ties. Despite their subjective nature these (often sub-conscious) value systems have operative meaning in controlling access of individuals to resources and decision making.
        We demonstrate these effects using a simple but sufficiently complex model with dynamical population size and network structure, where division of resources (prey) is carried out according to the principles of the Ultimatum Game. We focus on the 
        emerging proposer and responder thresholds under distinct social hierarchies and interaction networks and discuss them in relation to the extensive body of Ultimatum Game experiments conducted across a wide range of cultural contexts. 
      We observe the emergence of diverse sharing norms, ranging from unfair to highly generous, alongside the development of various social norms.
    \end{abstract}
 \maketitle
\section{Introduction}
The study of fairness or social cooperation has been explored for decades through the lens of games like the Prisoner's Dilemma and the Stag-Hunt game. Here, we choose a particularly simple game for this context, the Ultimatum Game (UG)~\cite{guth1982}. 
It is worth noting that zero-determinant strategies of the iterated Prisoner's Dilemma reduce to the UG once players notice that such strategies are played~\cite{press2012}. At that point, players must decide whether to accept the proposal or refuse to play.

In this work, we are interested in how status, and cultural means to "measure" status in terms of observables, such as energy levels or age of a player, bias the roles a player can take in cooperative games for resources. We are also interested how, e.g.~a minimal reproductive age and a dynamical population size affect the dynamics.

The UG, first studied experimentally by Güth et al.~\cite{guth1982}, involves two players with asymmetric roles---a "proposer" and a "responder"---and a reward to be divided. The proposer suggests how to split the reward, and the responder chooses whether to accept or reject the offer. If accepted, the reward is divided as proposed; if rejected, both players receive nothing.

According to the theoretical predictions, a rational and self-interested proposer will offer the smallest amount they expect the responder to accept. Likewise, a rational and self-interested responder will accept any positive offer and be indifferent between accepting or rejecting an offer of zero. Consequently, the subgame-perfect Nash equilibrium~\cite{gintis2000} prescribes that the proposer should offer the lowest possible unit of the reward (or zero), which the responder will then accept~\cite{camerer2011}. However, ample experimental evidence reveals that responders in Western, educated, industrialised, rich and democratic (W.E.I.R.D.) societies typically reject proposals at $10-30\%$ of the reward and proposers often offer $40-50\%$ (see e.g.~Oosterbeek et al.~\cite{oosterbeek2004}). While initially these thresholds were thought to be sharply defined and universally applicable, experimental evidence has unveiled various cultural and psychological biases that lead to either more greedy but sometimes also overly generous split proposals. For instance, Henrich et al.~\cite{henrich2005} conducted experiments in 15 small-scale societies around the globe and found a wide spectrum of arising strategies. From the "greedy" societies of the Quichua all the way to the Lamalera tribe, where offers frequently exceed $50\%,$ with proposers often giving more to others than keeping for themselves.

Despite significant cultural differences in fairness norms observed in experimental settings, the extent to which real-world actions deviate from the predictions of strict economic self-interest remains a topic of debate and has been examined from various theoretical perspectives (see e.g.~the review by Debove et al.~\cite{debove2016}, or more recent analysis by Akdeniz et al.~\cite{akdeniz2023}). Earlier models with role-alternating frameworks~\cite{rubinstein1982} suggest that what is learned through the repeated interactions might also translate to one-shot UG settings. Other models that have been investigated include evolutionary game-theoretical approaches with reputation-based explanations (see e.g.~\cite{nowak2000},\cite{chiang2008}), noise-based explanations~\cite{gale1995}, models with "empathy", where the agents are not allowed to ask for bigger splits than they would accept in the responder's role~\cite{page2002} and spite-based models~\cite{huck1999}. It has also been shown that the topology of the players' interaction network significantly influences the split levels that are commonly proposed or rejected (see e.g.~Sinatra et al.~\cite{sinatra2009} and Page et al.~\cite{page2000}). Interestingly, a small world transition takes place from relatively "fair" behaviour in a ring-shaped regular network topology to much more greedy splits in the small world, i.e.~for Poissonian networks~\cite{kuperman2008}. Furthermore, both~\cite{kuperman2008,page2000} present analytical invasion analysis that explores why non-zero rejection strategies can successfully stabilise within a structured population.

The majority of UG models assume that the roles of the players are dual, meaning they play both roles with equal probability. Consequently, when all individuals adopt the same strategy, the average payoff from successful interactions equals half the total reward. This inherent symmetry means the average total reward remains unaffected as long as the rejection threshold is below the proposal threshold.

However, this ceases to be true for societies that are not strictly egalitarian (i.e.~societies that do not really exist) but have notions of status that regulate the chances of an individual to play one or the other role in UG-like situations. Han et al.~\cite{han2018} experimentally highlight the difference between dual and single-role distribution by conducting experiments with different role assignment and different population structure (well-mixed vs.~fixed) and show that playing both roles (dual-role) promotes rational sharing (low offers), whereas single-role distribution leads to fair offers (approximately $45\%$).

Considering the inherent hierarchy between the players may therefore be an important component for studying fairness. Some authors have investigated unequal role distribution (see e.g.~\cite{li2013,wu2013,debove2015,killingback2001,snellman2019}).

For instance, Li et al.~\cite{li2013} look at degree-based role allocation and show fairer behaviour when low-degree players act more often as proposers. Wu et al.~\cite{wu2013} also look at unequal role distribution through role degree assignment on scale-free network, with their update mechanism the thresholds reach the highest values of $0.2$ if the proposers are played by low-degree nodes.
Debove et al.~\cite{debove2015} investigate the model with the possibility of switching partners at various cost levels and show that, for low costs, fairness can be achieved despite intrinsic unequal role distributions.
Killingback et al.~\cite{killingback2001} investigate a Collaborator's Dilemma similar to the UG. In their simulations, they assign to each of the players a dominance rank. During the game, the player with a higher dominance rank plays as the proposer and those with a lower rank play as the responder. If the ranks in the pair are the same, the game is not played. They observe different behaviour for different rank ranges, payoff functions, and different structures (fully connected network structure vs.~lattice).

Another group of interesting models deals with co-evolutionary dynamics of link adaptation/rewiring (see e.g.~\cite{gao2011,miyaji2013,deng2011,deng2021_link}). Gao et al.~\cite{gao2011} look at a model where players can rewire from those who are not offering them enough to make a deal. This is shown to promote fairness. 

Deng et al.~\cite{deng2011} examine a co-evolutionary model of the UG, focusing on the rewiring properties across different network topologies, and demonstrate that these dynamics promote fairness. Miyaji et al.~\cite{miyaji2013}, building on a similar framework, propose exploring additional co-evolutionary scenarios to identify which foster fairness and which do not. For example, allowing dissatisfied responders to relink results in very generous offers, whereas relinking by dissatisfied proposers encourages greedy strategies.

Ultimately, understanding the emergence of typical thresholds in accepting or rejecting offer (which may be multi-causal) remains a non-trivial task. Inter-personal behaviours, including proposing and rejecting strategies in UG-like situations, cannot be cleanly separated from other types of social interactions between individuals. 
In this paper, we study an evolutionary game setting in which players cooperate in pairs to receive nutritional energy, reproduce, and die of starvation or age. The division of the resource happens along  the lines of UG. We have two main goals in mind. First, we want to implement a "realistic" setting with dynamic environmental conditions, where players age and die, and only the segment of society endowed with sufficient energy can reproduce 
successfully within a reproductive age window. 

Second, we want to examine the role of "status" within the population, which determines role assignment in the game. In our model, the hierarchy is value-based, with subjective values unique to each individual. As a result, the structure is flexible and context-dependent, shaped by the values of interacting players rather than being rigid or universal. 

For instance, we want to explore how age and energy---observable traits that function as values in determining the relative status between two players---compete within the population.

We believe this framework is sufficiently complex to examine the co-evolution of status, which dictates role assignments in the game, alongside the values and attributes that define a player's status, and the development of offer and rejection threshold strategies.

Further, we are interested in how social structure, the minimum energy and reproductive age requirements impact population dynamics.

The paper is organised as follows: Section~\ref{sec:model} introduces the implemented model, followed by Section~\ref{sec:results}, where we present our findings. Section~\ref{sec:discussion} provides a discussion of the results and Section~\ref{sec:conclusions} concludes the paper.

\section{Model}\label{sec:model}
We consider a dynamically changing population with initially $N$ players connected either through a fully connected network or ring topology with a degree $k.$
In each round, players play multiple UGs with their neighbours from the social network and accumulate payoffs.
Every player $i$ is characterised by a vector of properties:
\[(a_i,o_i,v^i,p^i), \quad \text{for }i \in \{1,2,  \dots, N\},\]
where $a_i \in [0,1]$ is the accept threshold,  $o_i \in [0,1]$ is the offer threshold, $v^i \in [0,1]^V:\sum_{k=1}^Vv_k^i=1,$ is the vector with $V$ values, and vector $p^i$ includes attributes such as age, energy, and other observable features that may enter in determining the perception of status. The offer threshold defines the amount that the proposer offers to the responder. This offer is accepted if it 
exceeds the responder's accept threshold. In summary, in a game with reward $R$ between proposer $p$ with offer threshold $o_p$ and responder $r$ with accept threshold $a_r$ their respective payoffs, $\Pi_p$ and $\Pi_r$, are:
\begin{equation}
\begin{aligned}
\Pi_p&= \begin{cases}
(1-o_p)\cdot R &\text{if } o_p> a_r,\\
0 &\text{if } o_p\leq a_r,
\end{cases}\\
\Pi_r&= \begin{cases}
o_p \cdot R \quad \quad &\hspace{0.2cm}\quad \text{if } o_p> a_r,\\
0 \quad \quad&\hspace{0.2cm}\quad\text{if } o_p\leq a_r,
\end{cases}
\end{aligned}
\label{eq:payoff_description}
\end{equation}

The age of the player increases with every round and energy intake is also coupled to the games.  Every round each player randomly chooses a partner from the whole population in the case of fully connected network, or among their neighbours in the structured case, and they subsequently play a game. 
\paragraph*{\textbf{Values and Role Allocation.}}
The decision who plays the role of a proposer and responder depends on the relative status of the players in the pair. Each of the players computes their status and the status of the other player \textit{according to their values}. Values of player $i$ are determined by vector of weights $v^i \in [0,1]^V$ where $\sum_{k=1}^V v_k^i =1$. 
$v_1^i$ determines the level of importance of age for player $i$, for instance, $v_1^i=0$ means they do not consider age to be important, $v_1^i=1$ means the age is the only important attribute for building their status and judging the status of others. In similar way, $v_2^i$ determines the level of importance of energy, $v_3^i$ if someone's thresholds are in the "right" order (the offer threshold is higher than the accept threshold), $v_4^i$ the importance of being a \textit{good proposer} (number of accepted games played as a proposer divided by age), $v_5^i$ of being a \textit{good acceptor} (number of accepted games played as a responder divided by age),
and $v_6^i$ is the importance of the number of children (relative to their age).

Every player plays at least once per round, provided they have at least one neighbour. If they do not have any neighbour they do not interact in any games (and if this does not change, they will eventually die).

First, players in each pair have to decide who assumes which role. When players $i$ and $j$ play a game, they both assess their status based on their observable features and values.  
For each value category $k$, player $i$ determines whether they rank
"higher," "lower," or "equal" to the other player. For example, in the case of Age value, if player $i$ is older, then $q_1^i = 1$; if younger, $q_1^i = 0$; and if they are the same age, $q_1^i = \frac{1}{2}.$  
Finally, the self-esteem $s_i$($s_j$) of the $i$-th ($j$-th) player is determined as a weighted sum of $q^i$ and their value weights
\begin{equation}
    \begin{aligned}
        s_{i}= \sum_{k=1}^V v_k^i\cdot q_k ^i, \quad \quad
        s_{j}= \sum_{k=1}^V v_k^j\cdot (1-q_k^i),
    \end{aligned}
\end{equation}
where for all $k \in V$, the components of $q^i$ satisfy ${q^i_k\in \{0,\frac{1}{2},1\}},$ depending on the attributes of the players. 
Finally, the probability that the player $i$ is a proposer when playing against player $j$ is given as: 
\begin{equation}
\begin{gathered}
    p_{ij}= \frac{s_i}{s_i+s_j}.\\
\end{gathered}   \label{eq:p_ij}
\end{equation}
After the probabilistic rule determines who is the proposer, the payoffs are distributed as described in~\eqref{eq:payoff_description}.
\paragraph*{\textbf{Reproduction and Death.}} After the game is finished, every pair with an accepted deal can reproduce with probability $p_{\rm repro} = 0.1$, provided that both players are within a specified age range (from $\rm Age_{\rm min}$ to $\rm Age_{\rm max}$) and have enough energy. Energy threshold for reproduction is an evolving trait that is specific to each player. Each parent can have maximum one child per round. 

The new child expands the current population and the social network. In the fully connected case, the child is linked to every living player. In the structured population, the child connects to both of its parents and additionally also to nodes from a mixture of the parents’ neighbours and the broader population (see the next section for details on different wiring scenarios). The values, strategy thresholds and other evolving traits are inherited from the parent $i$ with probability $p_{ij}$ and $1-p_{ij}$ from the player $j$ (see~\eqref{eq:p_ij}). The copying process is imperfect with the uniformly distributed errors $[-\varepsilon,\varepsilon].$ If the resulting value falls outside the allowed range, it is rounded to the nearest permissible value. The child's age is zero and energy is inherited from both parents. How much either of the parents gifts to the child from their energy is also an evolving trait. We will refer to this child energy investment strategy simply as child-share.
 Finally, players are removed from the network together with all their links, if either their energy drops below $\varepsilon_E\ll 1$ or they surpass the age limit ($\rm Age_{\rm limit}$). 
The energy level of player $i$ in round $n$, denoted as $E^{n}_i$, is defined as
\begin{equation}\label{eq:energy_dyn}
E^{n}_i= E^{n-1}_i\cdot d - E_{\rm cost} + \Pi_i^n , \end{equation}
where $E^{n-1}_i$ is the energy from the previous round, which decays at rate $d,$ $E_{\rm cost}$ is the life cost per round, and $\Pi_i^n$ is the accumulated payoff gained from the UGs played in round $n.$ The probability of splitting a reward (prey) in every game is given as $\frac{R_n}{C},$ where $R_n$ is the overall resource level before playing UG, $S_n$ is the amount of resources eaten during $n$-th round and $C$ is the environmental capacity. After each round the resource is replenished with the following rate
\begin{equation}
    R_{n+1}= (R_n-S_n) \left(1+\gamma \max\{1-\frac{R_n-S_n}{C},0\}\right),
\end{equation}
\color{black}
where $\gamma=\frac{1}{12}$ (one round represents $\Delta t=1$ month).

In the following section we specify the parameters and simulation setups that we used and present the simulation results.
\begin{figure*}
      \includegraphics[height=6.8cm,trim=5cm 3.7cm 5.8cm 3.8cm, clip]{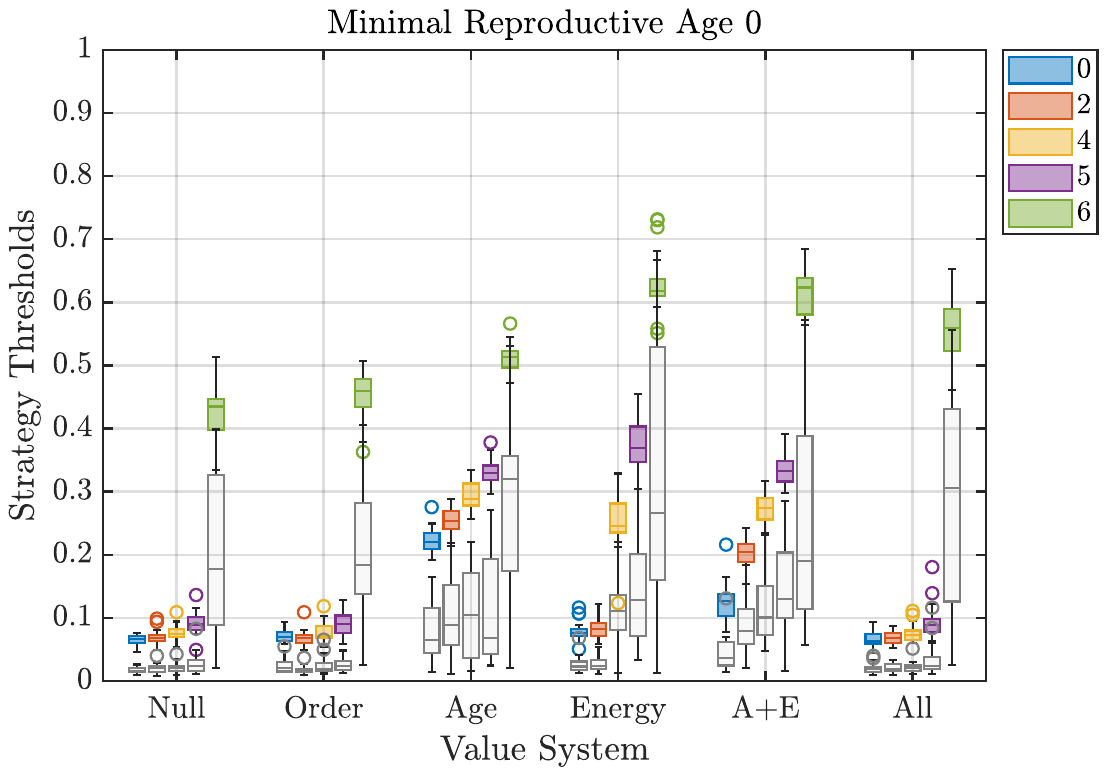}
      \includegraphics[height=6.8cm,trim=5.9cm 3.7cm 7.8cm 3.8cm, clip]{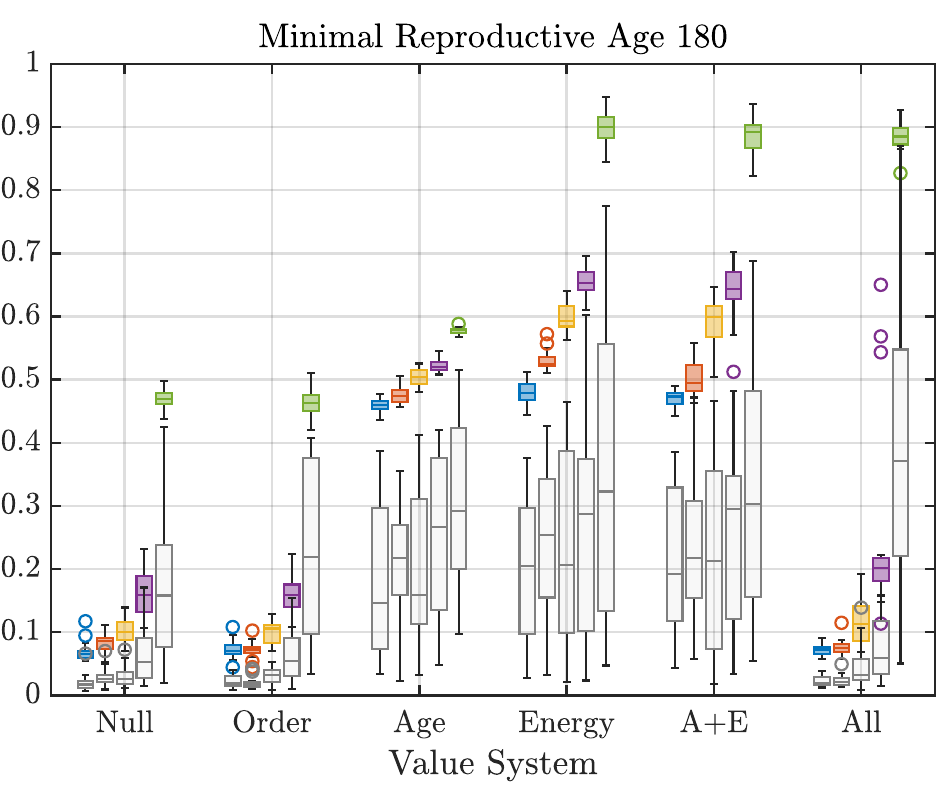}
   \caption{Evolved thresholds of offers (coloured) and rejections (grey) in simulations with a structured network and a minimal reproductive age of zero (left) and 180 month (right). Different colors represent various wiring configurations: blue (0) indicates that a child node is connected to both parents and the remaining six links are assigned randomly from the entire population. Red (2) represents simulations with two of the six remaining links selected from the parents' neighbours, and so on until the case where the child node is connected to parents and their neighbours only (6). Each box represents $20$ means, where each mean is calculated from the population distribution in a single run. The x-axis displays different value systems, while the y-axis shows boxplots of mean offer thresholds (coloured) and acceptance thresholds (grey).}
   \label{fig:results_sw}
\end{figure*}
\section{Results}\label{sec:results}
In this section, we analyse the simulation results of the model, exploring various configurations. We investigate evolved thresholds, social network topologies, and "cultures" under different reproductive regimes, value systems, and wiring scenarios. First, we present the results from a simple setup with a fully connected network, and then proceed to the main findings from a structured network. The simulations were conducted using the following parameters.

For each distinct configuration $20$ runs were performed. The number of players in the initialisation is $250.$  In the structured case, the initial topology of the network is a ring structure with degree $8.$  The length of the simulations is $2\times10^6$ time steps (one time-step represents one month), the age limit is set at $\rm Age_{\rm limit}=1000$ steps (approximately $83$ years). The minimal reproductive age is varied ($0$ vs.~$180$) and the maximal reproduction age is set to $\rm Age_{\rm max}=450.$ The reproduction probability is given as $0.1$ reminiscent of the human reproductive probabilities. The strategy thresholds, values and attributes are initialised on uniform intervals between the respective minimal and maximal limits (and in the case of values then normed to sum to 1). The initial energy of every player is set at $5.$ The maximum magnitude of the noise of the replication $\varepsilon$ is given as $0.005.$ Note that if $\varepsilon$ was set to zero, extinct strategies would have no chance of re-appearing. Conversely, setting the noise too high would lead to emergence of entirely random strategies and properties. The cost of life, $E_{\rm cost},$ is set as $1,$ the decay rate $d$ as $1-1/12$ (see Eq.~\eqref{eq:energy_dyn}). The average number of games played is two per round per player and the reward size is~$5.$ The probability of receiving the reward is proportional to the resource level (see previous section). The caring capacity is set as $10000$ and resources are initialised at their full capacity.

We analyse six different value systems:
\begin{enumerate}
    \item Null (no values, equal probability for either role)
    \vspace{-0.15cm}
     \item Order   \vspace{-0.15cm}
    \item Energy   \vspace{-0.15cm}
    \item Age   \vspace{-0.15cm}
    \item Age + Energy    \vspace{-0.15cm}
    \item Age + Energy + Order + Good Acceptor + Good Proposer + Reproduction per year 
\end{enumerate}
For conciseness, we may refer to a value system simply as a system when the meaning is clear from context. Similarly, we will refer to the Age+Energy system as "A+E" system, and to the $6th$ value system as "All" system.

\subsection*{\normalsize Fully Connected Network}
As a starting point, let us consider the case with a fully connected network and analyse how the simulations behave in the simplest, unstructured scenario. 

When the minimal reproductive age is set to zero, only populations in four value systems persist without going extinct: the Null system, the Order system, the Energy system (in only five out of twenty simulations), and the All system. In runs where the minimal reproductive age is increased to 180, populations with the Energy value system go extinct, while the other three systems continue to survive. In all non-extinct simulations, offer thresholds evolve to “rational” (i.e.~greedy) levels, between $3\%$ and $9\%,$ as is also predicted by theory. In the All value system, the strongest evolved value is consistently Order, comprising approximately $76\%$ to $90\%$ of the population's value's weight distribution. Additionally, in each run, $98\%-100\%$ of players have their strategy thresholds ordered such that the offer threshold is higher than the accept threshold. Thus, the Order value system is very similar to the Null system.

For zero minimal reproductive age with the Energy value system, the Gini index with respect to the energy distribution in the population ranges from $0.5$ to $0.54.$ For the other three value systems, it is lower---indicating a more equal energy distribution---ranging from $0.32$ to $0.4.$ Note that this still represents a relatively high level of energy inequality within the population, caused by the probabilistic assignment of roles in each round and the presence of very greedy ("imbalanced") thresholds, where proposers retain almost the entire reward for themselves.

When the minimal reproductive age is raised to $180,$ Gini index drops to values between $0.24$ and $0.31$ for the Null, Order and All value systems. 
Under the Energy value system and minimal reproductive age zero, the average age is the highest (approximately $150$ months) compared to other value systems (approximately $42$ months). Not surprisingly, when we raise the minimal reproductive age to $180$ months, the average age grows to values of approximately $200$ to $300$ months. 

The size of the child-share (i.e.~how much energy parents gift to their children) varies across populations in different scenarios. Whereas, with a minimal reproductive age of zero, parents in the Energy value system typically give, on average, $8\%$ of their energy to their children, in the other three value systems the contribution is considerably higher, at around $12-21\%.$ Similarly, in the case of minimal reproductive age $180$ the child-share is also higher, at around $14-30\%.$ 
\subsection*{\normalsize Ring Network}
In this subsection, we present simulation results for the structured network, initialised with a ring topology of degree $8$ (a realistic degree for human social networks).

We examine various scenarios for the formation of new links to child nodes. In all scenarios, each child is connected to both of their parents, with additional links added from the whole population. The primary difference between scenarios lies in how these extra links are allocated---some are fixed to connect to the neighbours of the parents, while others are assigned randomly to arbitrary individuals in the population.
In the scenario with fully random linking (labeled as $0$), in addition to two links connected to the parents, six nodes are randomly selected from the entire population, and the child links to them (note that among the six randomly chosen nodes, some may coincide with the parent(s); this is treated as a stochastic element, where not all children necessarily have eight distinct new links). In the second setup (denoted as $2$), two of the six links are chosen from the parents' neighbours, while the remaining four are assigned randomly. The third setup increases this number to four, the fourth setup to five, and in the final version (denoted as $6$), all six links are exclusively selected from the parents' neighbours (excluding the parents' nodes).

The main results are depicted in Figure~\ref{fig:results_sw}. Resulting strategy thresholds of proposing (coloured) and rejecting (grey) behaviour vary significantly across all six linking scenarios and two distinct setups with different values of the minimal reproductive age ($0$ vs.~$180$). Next, we will analyse this behaviour and also explore other evolved properties in detail.
\begin{figure}
    \centering
    \includegraphics[width=\linewidth,trim={1.85cm 1.05cm 2.5cm 1.2cm},clip]{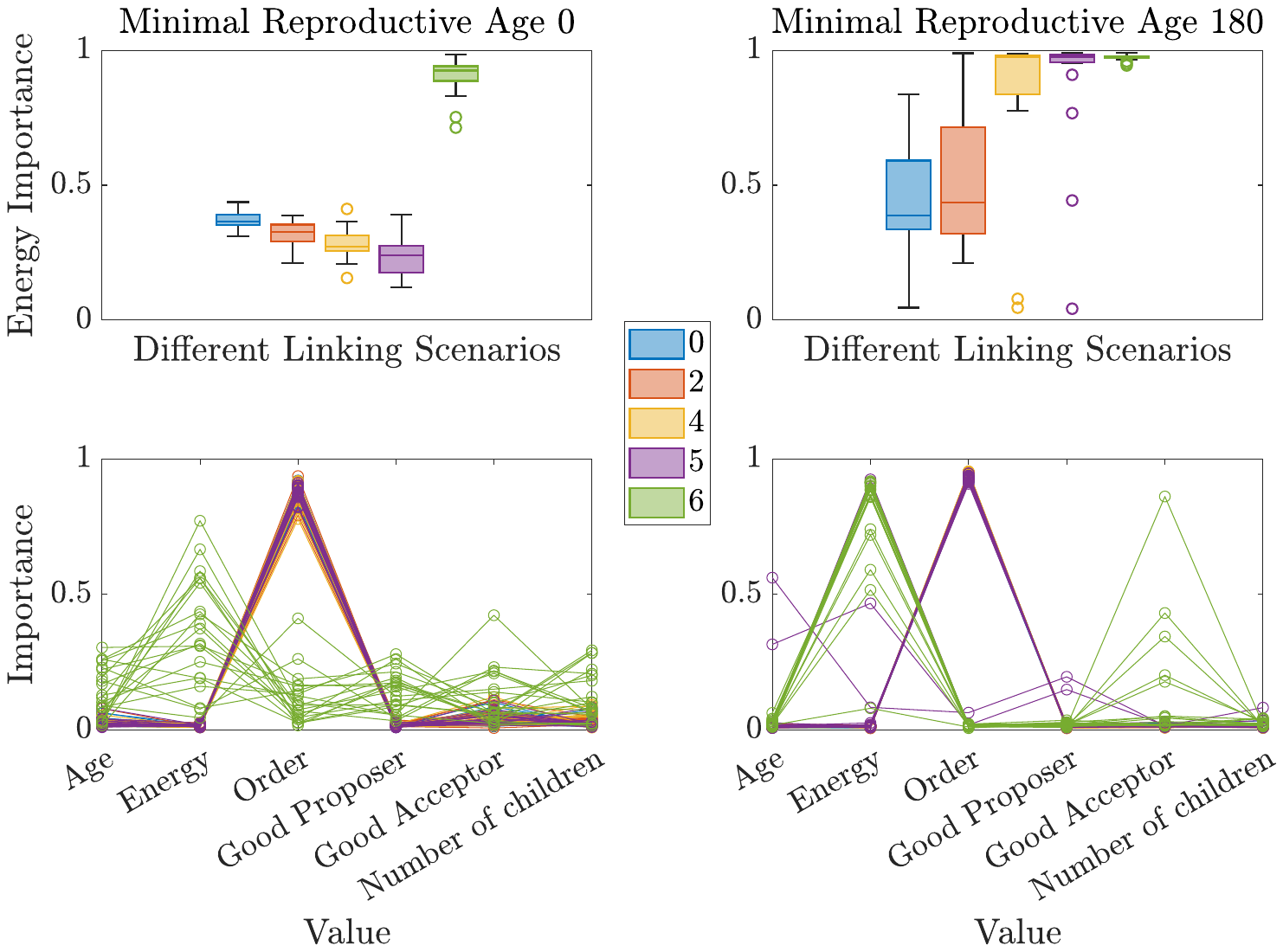}
    \caption{Evolved value systems for different settings. The upper plots show energy importance in the mixed systems of Age + Energy. The plots below show results for the All system values. On the left, we show results for minimal reproductive age of zero, on the right for age $180.$ Different colors refer to distinct wiring scenarios (see the legend).}
    \label{fig:values_sw}
\end{figure}
\begin{figure*}
    \centering  
 \includegraphics[width=0.4\linewidth,trim={4.5cm 8.5cm 4.1cm 9.5cm},clip]{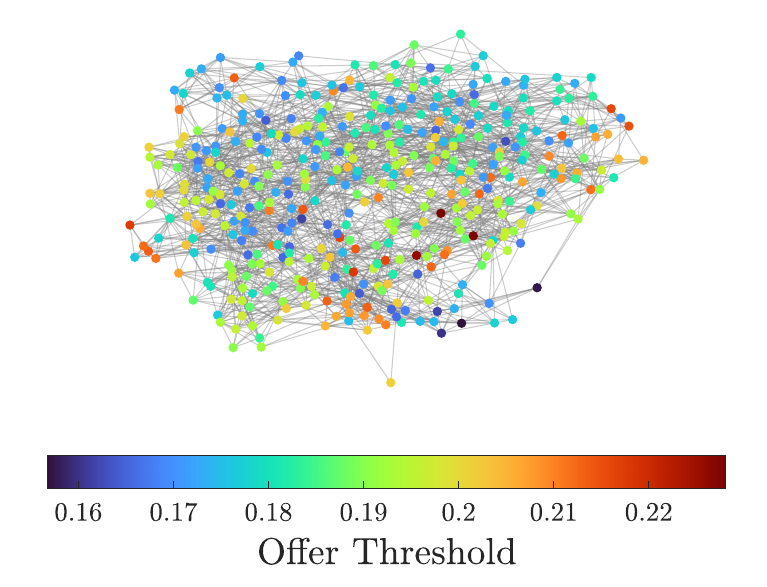} \hspace{1cm}
\includegraphics[width=0.4\linewidth,trim={4.5cm 8.5cm 4.1cm 9.5cm},clip]{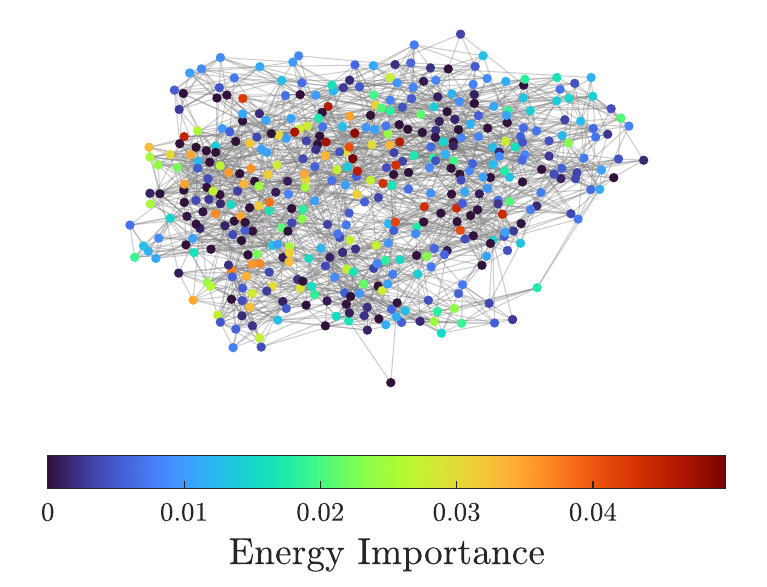}
\includegraphics[width=0.4\linewidth,trim={4.5cm 8.5cm 4.1cm 9.5cm},clip]{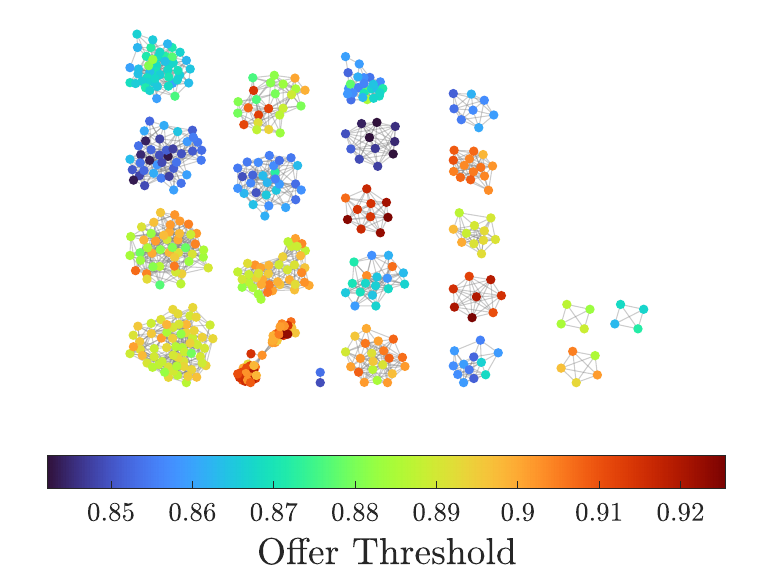} \hspace{1cm}
\includegraphics[width=0.4\linewidth,trim={4.5cm 8.5cm 4.1cm 9.5cm},clip]{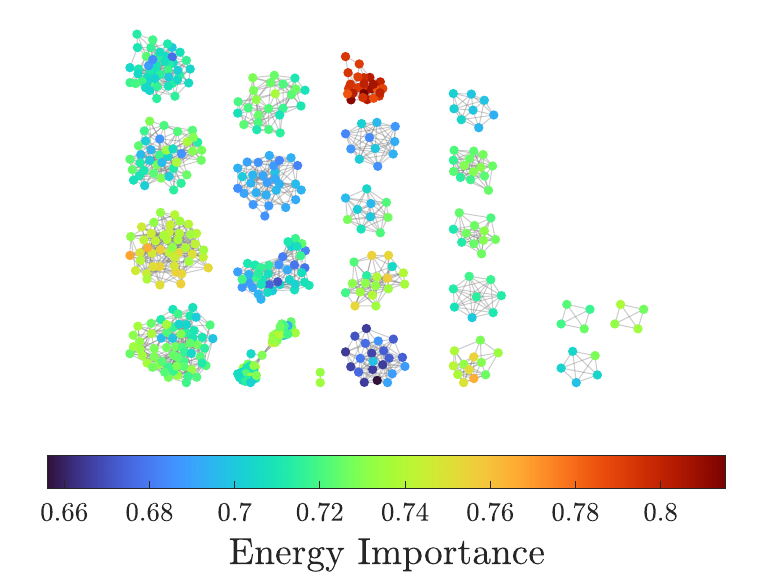}
    \caption{Network structure of the case with All value system and a minimal reproductive age $180.$ Different colors represent different offer thresholds in society (left) and different levels of energy value strength (right). The plots above have one link connected to a random individual in the entire population and the rest to neighbours (purple wiring (5)) and the plots below represent the wiring case where all links come form parents' neighbours (green wiring (6)).}
    \label{fig:sw_network_structure}
\end{figure*}

\paragraph*{\textbf{Minimal Reproductive Age Zero.}}
First, let us examine the results where all agents reproduce from birth until 450 months. Let us start with the scenarios involving (partially) random wiring, where at least one of the child's links is assigned to a randomly selected player rather than being exclusively inherited from the parents' neighbours---see the box plots labeled 0 to 5 in Figure~\ref{fig:results_sw} (thus excluding the green ones). In this case, the offer thresholds remain low, around $5-15\%$ for the Null, Order, and All systems. Note that under these conditions, the population in All value system primarily adopts Order as the dominant value (see the bottom-left plot in Figure~\ref{fig:values_sw}).
Whereas, with a minimal reproductive age of zero, parents in the Energy value system typically give, on average, 8

\begin{figure*}
       \begin{overpic}[width=0.325\linewidth,trim={7.5cm 5cm 8.2cm 5.5cm},clip]{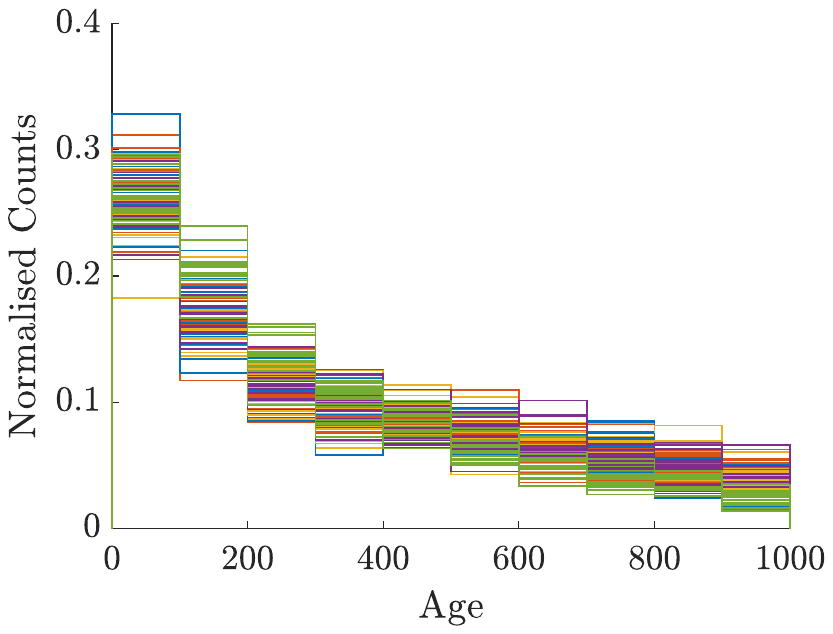} 
            \put(-1,75){\small a)}
        \end{overpic}    
   \begin{overpic}[width=0.325\linewidth,trim={7.5cm 5cm 8.5cm 5.45cm},clip]{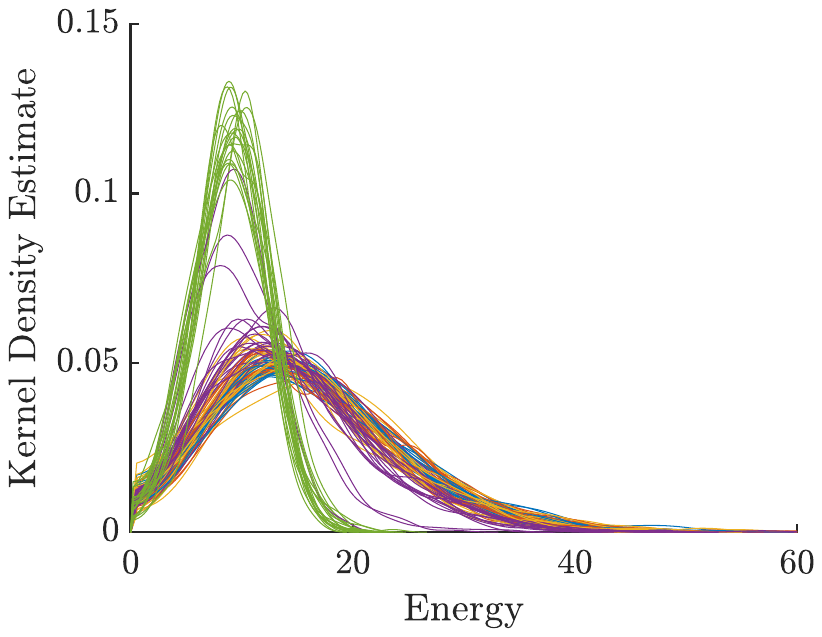}     
             \put(-1,75){\small b)}
        \end{overpic}
     \begin{overpic}[width=0.325\linewidth,trim={7.5cm 5cm 8.5cm 5.45cm},clip]{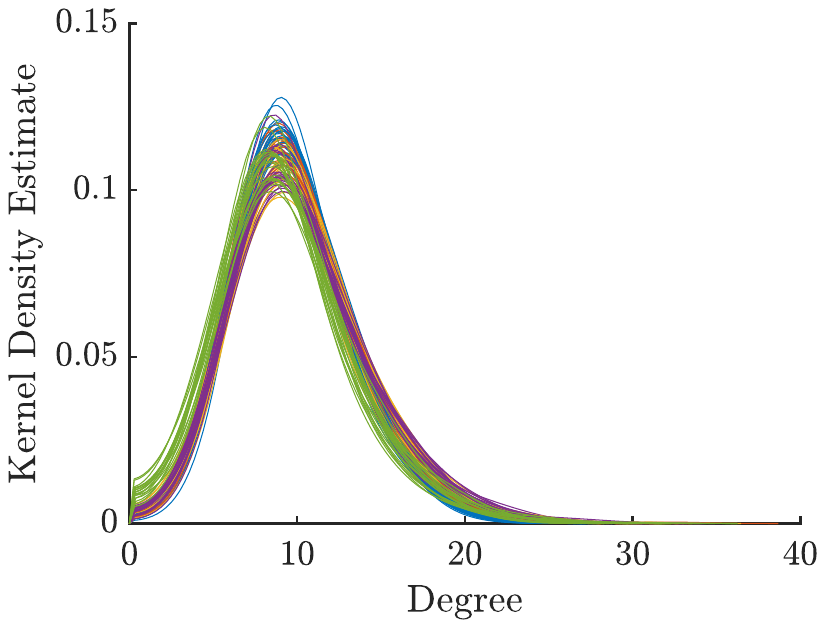}
            \put(-1,75){\small c)}
        \end{overpic}
     \caption{Age (a), energy (b), and degree (c) probability density estimates are shown for different wiring scenarios in the setups with a minimal reproductive age of $180$ and the All value system. A realistic age distribution is observed in panel (a). The energy distribution (b) differs noticeably in the green wiring scenario (6), with a much narrower distribution and lower average energy values compared to other scenarios. The degree distribution (c) is quite similar across wiring scenarios, with an average degree of approximately $9.9$.} \label{fig:distributions}
 \end{figure*}

In contrast, when connections are based purely on parents' neighbours (represented by the green box-plots in Figure~\ref{fig:results_sw}), the results are strikingly different. Offer thresholds often surpass $50\%$ across all value systems, except for the Null and the Order systems---where they still remain noticeably higher than in the other wiring scenarios.
The network structure differs strongly from the other scenarios---network becomes highly fragmented into small, isolated components (or \textit{clans}) each with its own distinct "culture" (in terms of values) and offer thresholds (see Figure~\ref{fig:sw_network_structure} for a similar situation result illustrating the network structure in the case of minimal reproductive age $180$). In this case, depending on the trial, we observe the formation of $4$ to $17$ isolated network components. Furthermore, values vary widely and there is no clear ranking of values in All value system, whereas in A+E system the Energy value is the leading value (see the green coloured scenarios in the left panels of Figure~\ref{fig:values_sw}).
\paragraph*{\textbf{Minimal Reproductive Age $180.$}} When the minimal reproductive age is increased to 180, the evolved thresholds differ notably from those in zero minimal reproductive age, across all value systems except for the Null and the Order system (and for some scenarios in the All value system) where the difference is relatively small. In the other value systems, thresholds are higher compared to scenarios where players reproduce from age zero. For both Energy and Age and their mixture (E+A) value systems for purely random wiring (see blue box-plots in Figure~\ref{fig:results_sw}), the offer thresholds reach values around $45-50\%.$ With growing number of child links coming from neighbours of the parents, the offers are rising, reaching values of about $45-70\%,$ with Age system thresholds being lower on average. Lastly, when all links come from the neighbours of parents (green box-plots in Figure~\ref{fig:results_sw}) the offers in the Energy and E+A system become very generous reaching approximately $90\%$ and in the
Age system the offers are just below $60\%.$
In contrast, in the All values system with (partially) random wiring, offer thresholds remain relatively low, similar to the levels observed in the Order and Null systems, ranging between approximately $5\%$ and $25\%,$ with the exception of few outliers in the purple regime (5 links connected to random players). In these scenarios, the dominant value is Order (see Figure~\ref{fig:values_sw}). However, when all links are assigned exclusively to the parents' neighbours, the offer thresholds become very generous again, with Energy (and, in a few cases, the Good Acceptor) emerging as the dominant value (again, see Figure~\ref{fig:values_sw}). In the A+E system, the leading value is primarily Age for random wiring (wiring 0 and 2), but then it shifts to Energy in the other scenarios. In the setup with minimal reproductive age of 180, we observe creation of $16$ to $27$ network components, which is significantly more than $4$ to $17$ components in the scenario of minimal reproductive age zero. Figure~\ref{fig:sw_network_structure} illustrates one example of such a fragmented network, along with a case from the linking scenario $5.$ Finally, in Figure~\ref{fig:distributions} we show the age, energy and degree distribution of the simulations with minimal reproductive age $180$ and All value system.
 
\paragraph*{\textbf{(In)equality, Average Age and Child-Share.}}
\begin{figure*}
        \includegraphics[height=6cm,trim={4.7cm 3.9cm 5.9cm 4.1cm},clip]{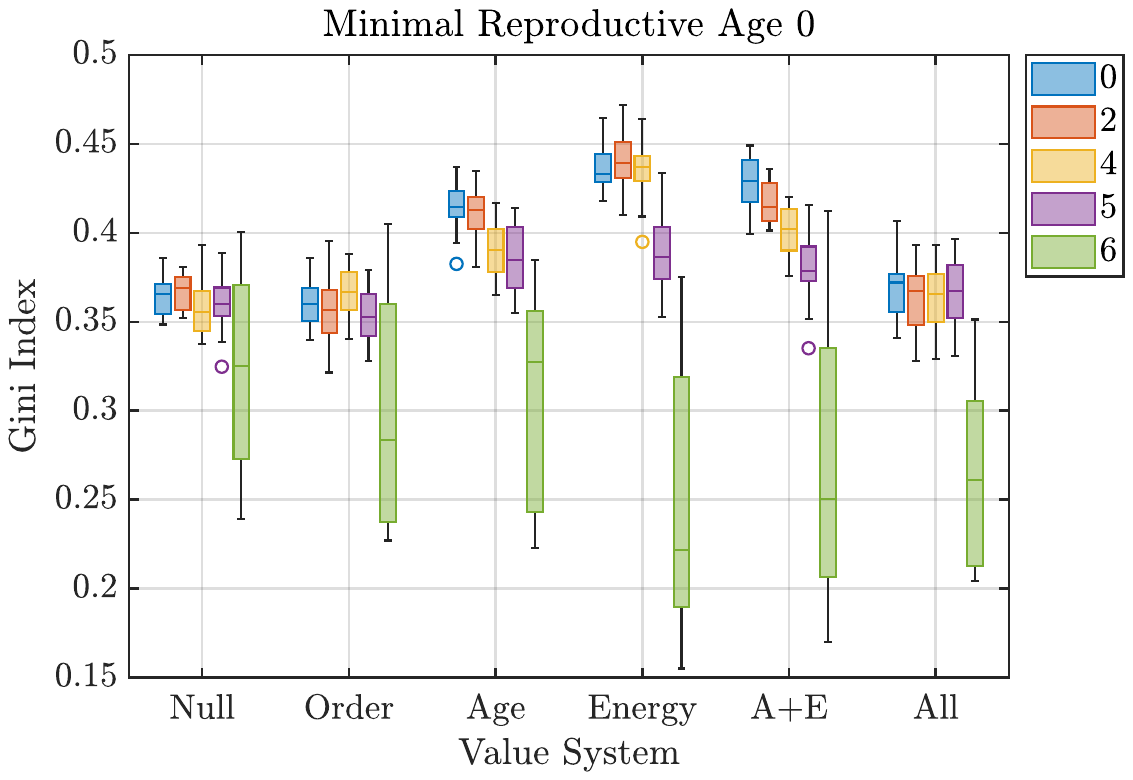}
        \includegraphics[height=6cm,trim={4.7cm 3.9cm 7.79cm 4.1cm},clip]{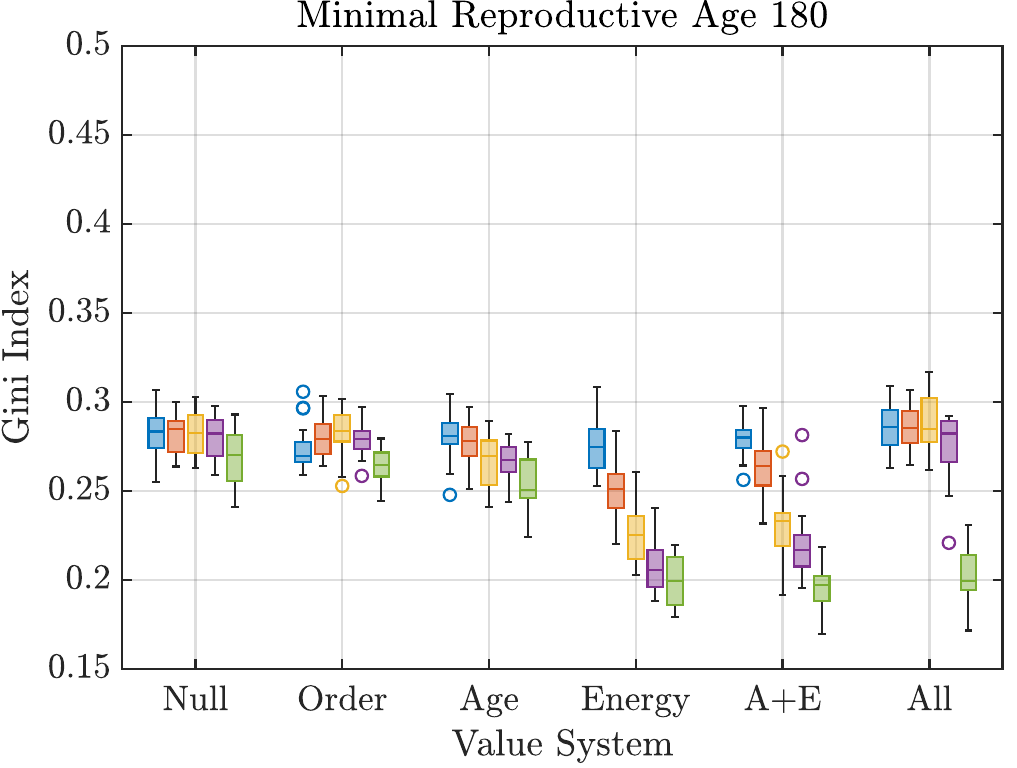}
    \caption{Gini index of the energy distribution in the evolved populations under different wiring scenarios and value systems. On the left, the results for a minimal reproductive age of zero are shown, and on the right, those for a minimal reproductive age of 180. Different colors represent various wiring scenarios, while the x-axis displays different value systems.}
    \label{fig:gini_idx}
\end{figure*}

Let us now examine the inequality levels in each scenario, which we measure using the Gini index of energy distribution. The Gini index ranges from zero, indicating perfect equality (everyone has the same energy), to one, which reflects total inequality (one person has all the energy, and everyone else has zero). As shown in Figure~\ref{fig:gini_idx}, the Gini indices in all scenarios range from approximately  $0.15$ to $0.47.$ Despite the Null system having an equal distribution of roles, the Gini index is relatively high. This arises due to the probabilistic nature of player-pairing for each round and a heterogeneous degree distribution, combined with random allocation of roles (and potentially unequal reward splits), varying numbers of children, and other factors. The (partially) random wiring scenarios have generally higher Gini indices under the minimal reproductive age of zero compared to the age 180. 
Societies constructed with purely neighbour based wiring (green, label 6), have the most equal outcomes with the lowest Gini indices across all value systems. Equal role distribution has the highest level of equality under $50\%$ reward splitting, a condition nearly achieved in all simulations of Null and Order-based systems.

In the green scenario where Energy is the leading value and there is a hierarchical role allocation (i.e.~Energy, A+E and All value system of minimal reproductive age $180$), inequality is mitigated by awarding the energetically inferior player a larger share of the reward, resulting in lower Gini index values than in any other scenarios (see Figure~\ref{fig:gini_idx} and also Figure~\ref{fig:distributions} with the energy distribution).
\begin{figure*}
    \centering
                \includegraphics[height=6cm,trim={4.9cm 3.9cm 5.9cm 4.1cm},clip]{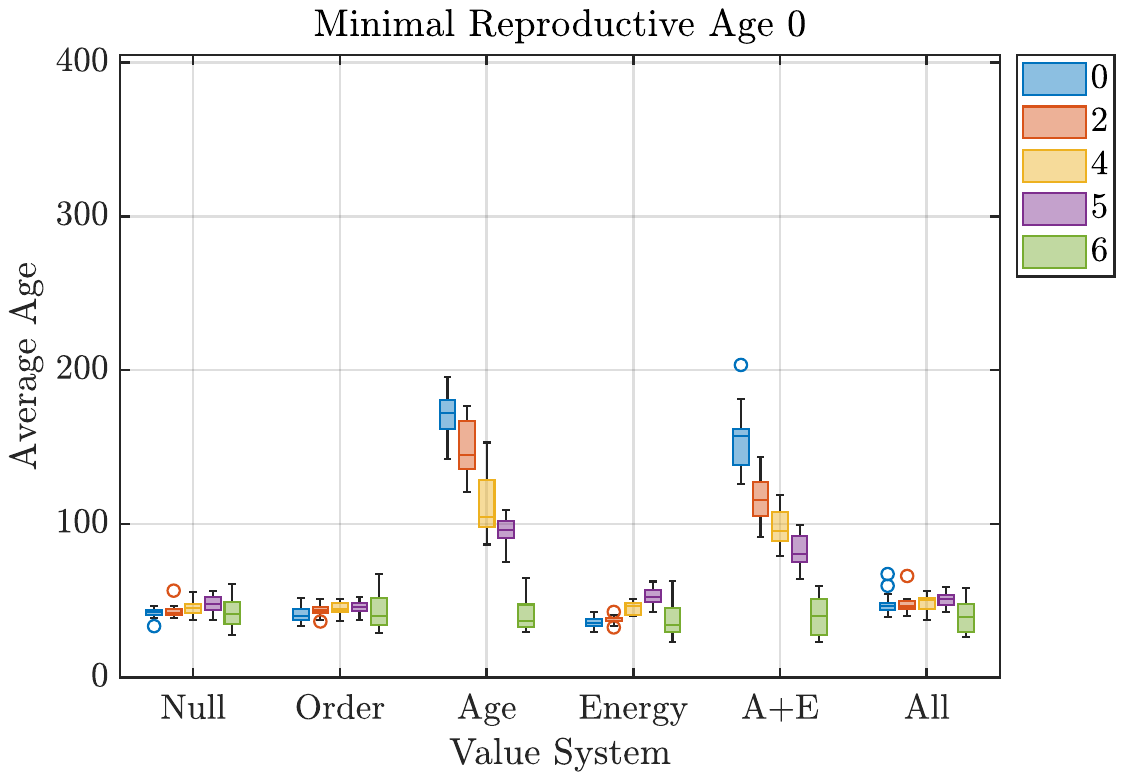}\quad
                \includegraphics[height=6cm,trim={4.9cm 3.9cm 7.79cm 4.1cm},clip]{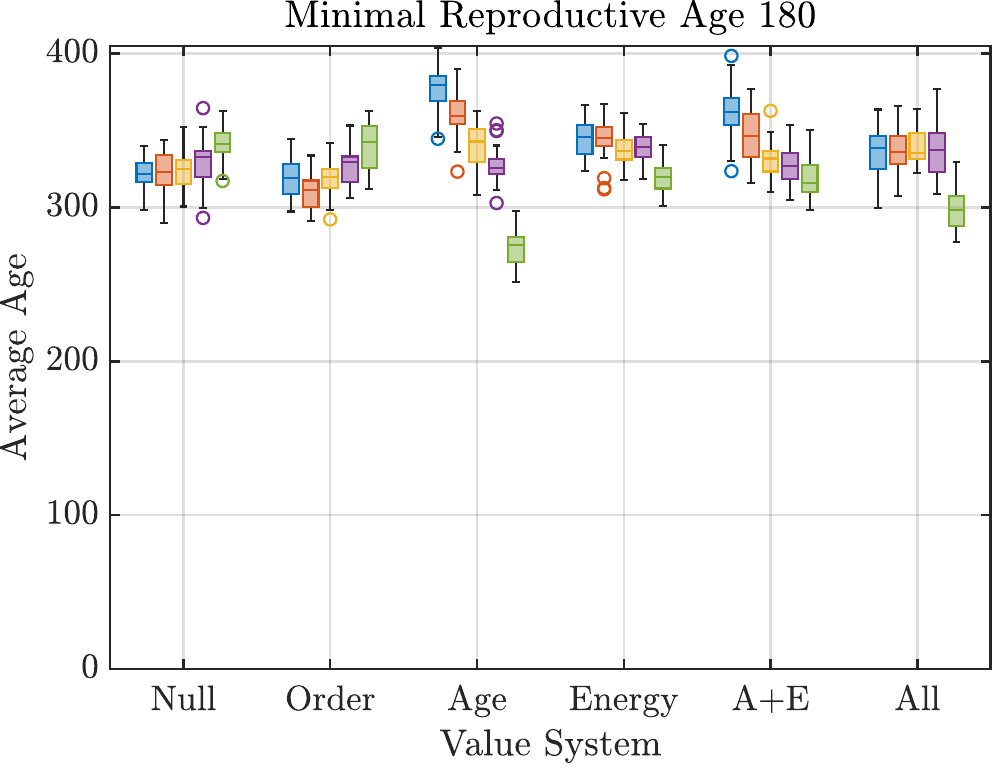}
    \caption{Average age of the living population under different wiring scenarios, value systems and reproductive windows. On the left (right) the minimal reproductive age zero (180). Different colours represent different wiring scenarios (see the legend) and on x-axis there are different value systems.}
    \label{fig:age}
\end{figure*}
\begin{figure*}
    \centering
\includegraphics[height=6cm,trim={4.7cm 3.9cm 5.9cm 4.1cm},clip]{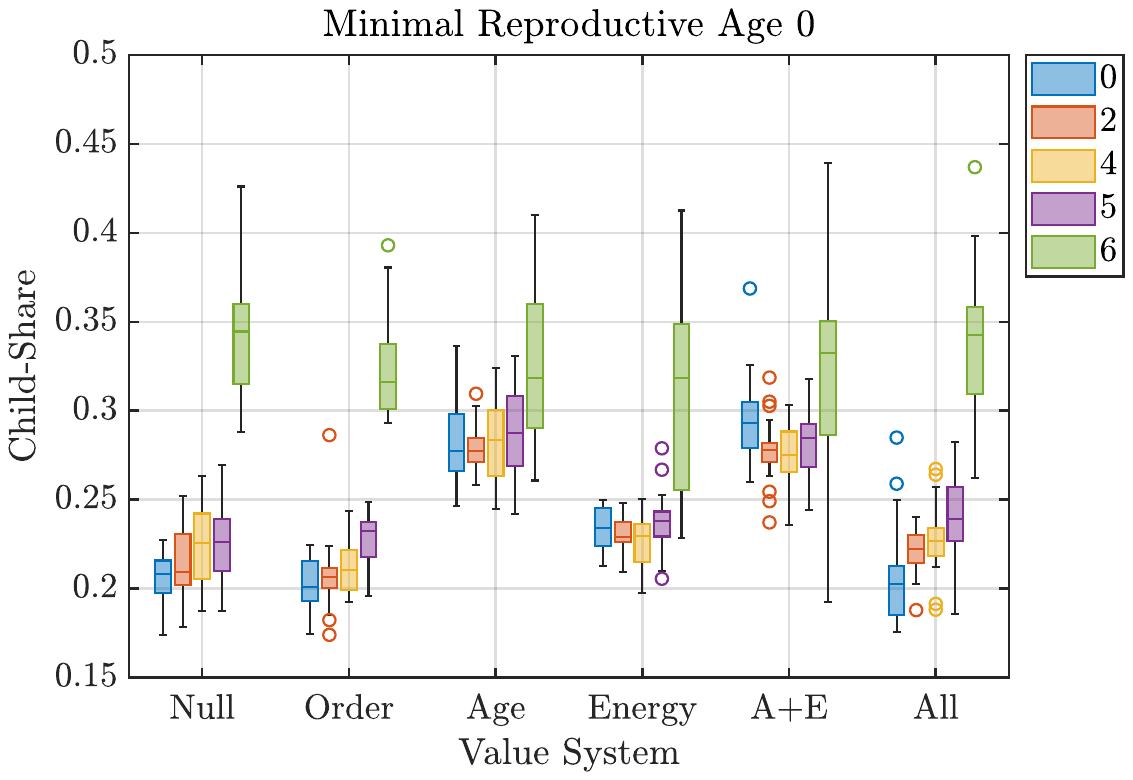}\quad
       \includegraphics[height=6cm,trim={4.7cm 3.9cm 7.79cm 4.1cm},clip]{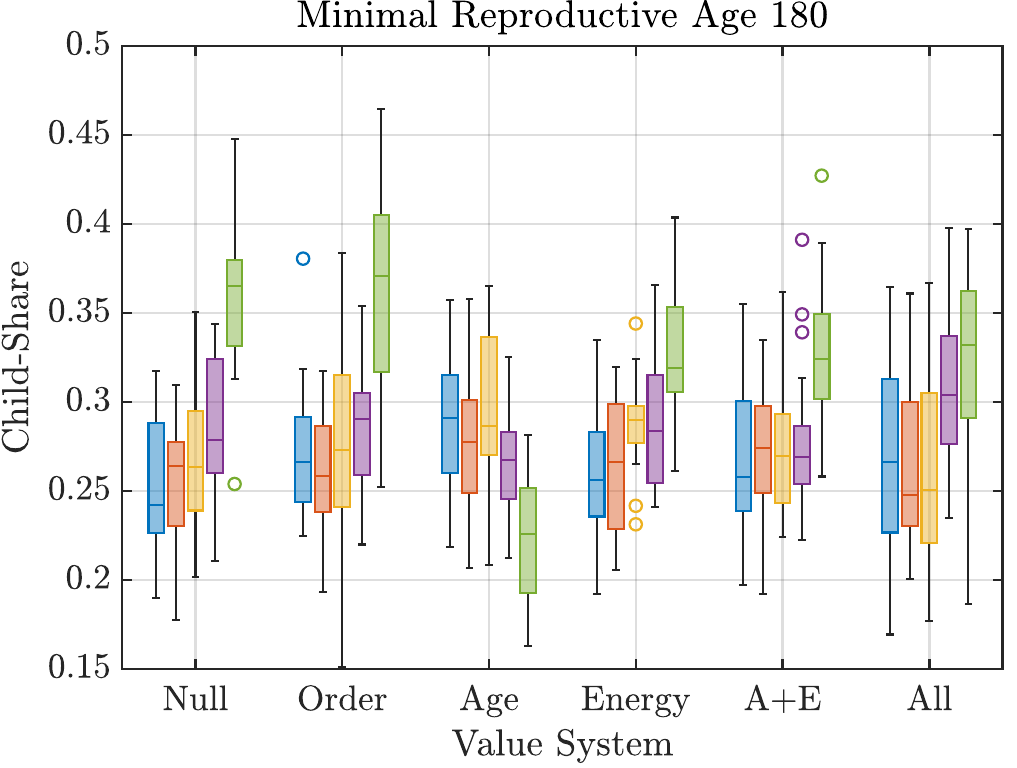}
    \caption{Average child-share in the evolved populations under different wiring mechanisms (see the legend), values systems (see the x-axis) and two setups with the minimal reproductive age  of zero ($180$) on the left (right).}
    \label{fig:child_share}
\end{figure*}

Comparing the scenarios with a minimal reproductive age of zero versus 180 months, we observe differences in both the population's average age (of living individuals) and the child-share strategy. When the minimal reproductive age is zero, the average age remains relatively low, around $20-70$ months across most value systems, except for the Age and A+E systems, where it increases to approximately $70–200$ months for (partially) random wiring. In contrast, with a minimal reproductive age of 180 months, the average age rises significantly, reaching approximately $250–400$ months across all systems. In Figure~\ref{fig:distributions} we show age distributions for various scenarios of minimal reproductive age $180$ and All value system.

The child-share strategy varies considerably under both reproductive regimes (in terms of minimal reproductive age), ranging from $0.15$ to $0.47.$ Note, that if both parents contribute $33\%$ of their energy reserves, all three individuals (both parents and the child) end up with equal energy levels. However, while parents may gain energy in other games played during the round, they also lose energy at the end of it, whereas children, during their first time step (as a rule), do not. In the green wiring scenario, the child-share is the highest across all cases except in the Age value system with a minimum reproductive age of 180 months, where it drops to a median value of 0.23.
\paragraph*{\textbf{Similarity of the values.}}
\begin{figure*}
\includegraphics[width=8.5cm,trim={4cm 8.2cm 7.8cm 6.35cm},clip]{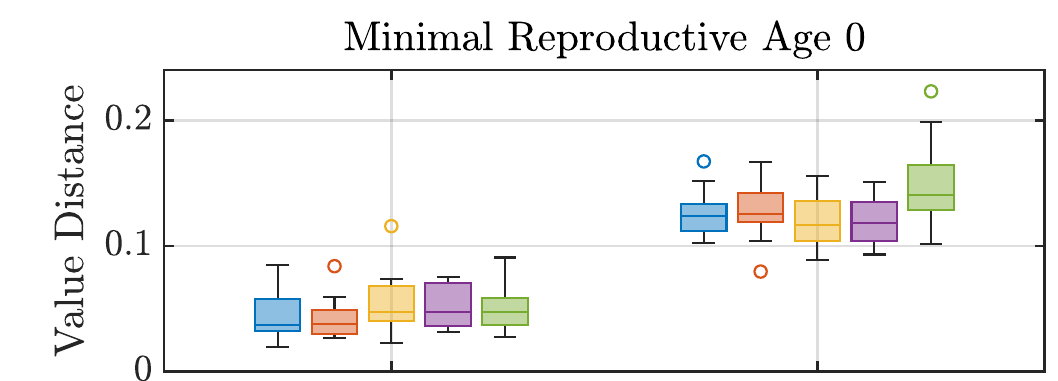}\hspace{0.6cm}
\includegraphics[width=8.5cm,trim={4cm 8.2cm 7.8cm 6.35cm},clip]{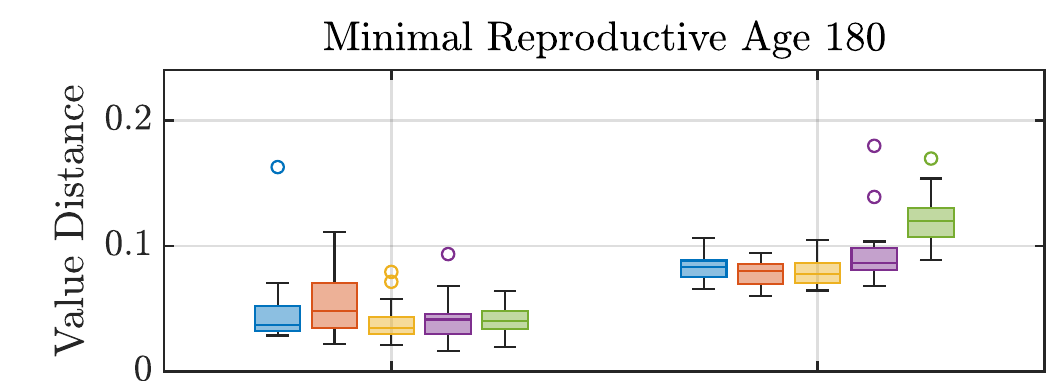}
\hspace{-0.5cm}
\includegraphics[width=9.4cm,trim={4cm 6.2cm 5.8cm 7.35cm},clip]{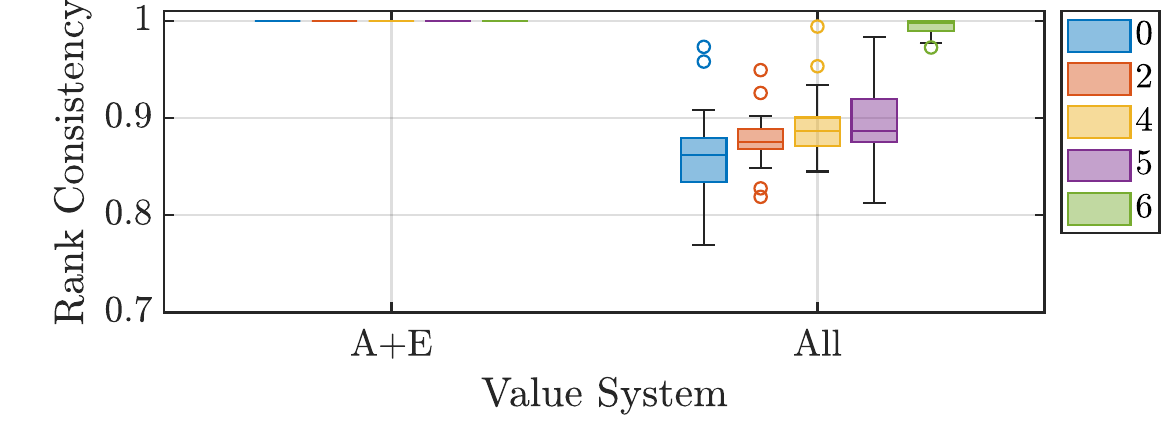}
   \hspace{-0.4cm}    \includegraphics[width=8.5cm,trim={4cm 6.2cm 7.8cm 7.35cm},clip]{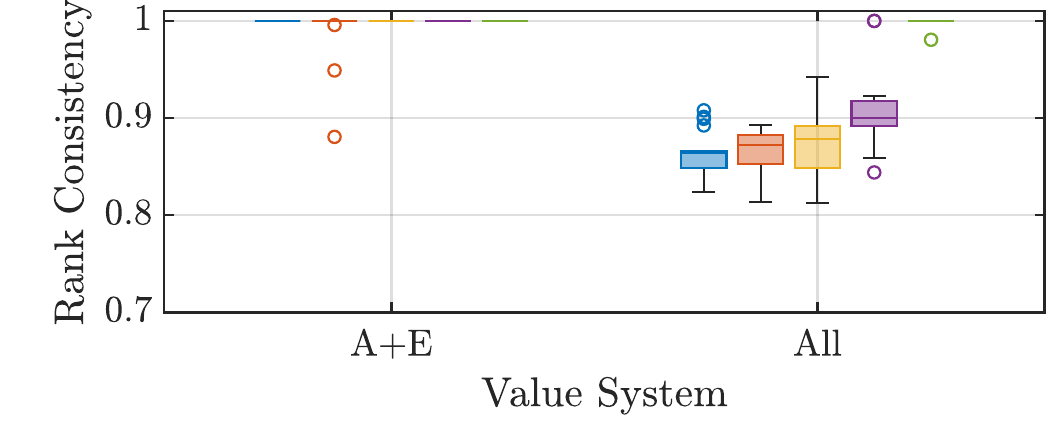}
    \caption{On the plots above, average $l_1$ distance between individual players' value vectors. On the plots below, rank consistency. The left (right) panels shows results for various wiring mechanisms (see the legend) with a minimal reproductive age of zero (180). We show the results for two value systems (see the x-axis).}
    \label{fig:values_sim}
\end{figure*}
\begin{figure}
    \centering
\hspace{-0.2cm}\includegraphics[height=4.1cm,trim={5.2cm 5.8cm 6cm 5.5cm},clip]{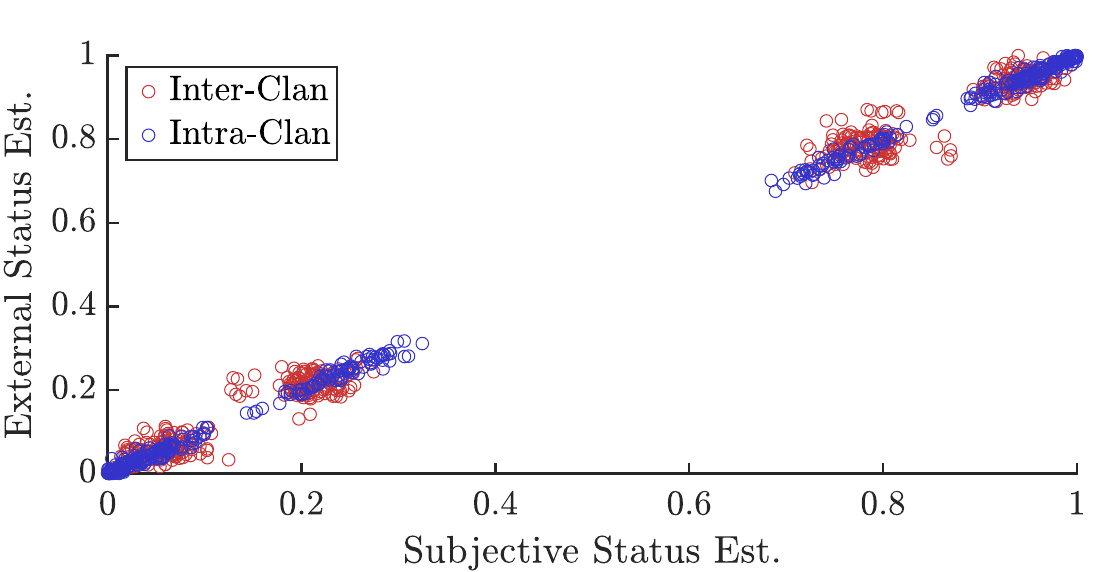}
    \caption{Subjective vs.~external status estimates of all players in the last round of a simulation with the All value system, with a minimal reproductive age $180$ and the green wiring scenario (6). Each circle represents a pair of players in the society. The first player’s estimate of their status relative to the second player is plotted against the second player’s estimate of the first. If both assigned identical value weights, their estimates would align along the identity line. However, due to individual differences in value weighting, this alignment rarely occurs. Blue points represent estimates from co-players within the same network component, while red points are from different components, indicating stronger alignment within clans. The highest average value weights in this trial are Energy with $72\%$ importance and Good Acceptor with $20\%.$ Thus, the four clusters represent combinations of relative attributes: (1) the first player is energetically poorer and worse acceptor, (2) poorer and better acceptor, (3) richer and worse acceptor, (4) richer and better acceptor than the co-player.}
   \label{fig:cigar}
\end{figure}
Next, we examine the \textit{similarity} of values. To do so, we compute the average $l_1$ distance between the players' value vectors, including those who are not connected in the network. The results for the A+E system and the All system are shown in the upper panels of Figure~\ref{fig:values_sim}. For a \textit{naive} population with random values, the average absolute distance between players' values is approximately 
$0.54$ when the system has two values and around $0.65$ when there are six values. Thus, results indicate that the values in the population become significantly more aligned compared to the naive case. 
Additionally, we analyse rank consistency (see bottom panels of Figure~\ref{fig:values_sim}), i.e.~how often players in a pair agree on who has the higher status. This is calculated only between connected players in a network. We see a strong agreement in their rankings ($76\%-100\%$). Lastly, in Figure~\ref{fig:cigar}, we present one simulation result from All system with minimal reproductive age 180 and a wiring scenario where children connect only to parents' neighbours (green regime). It illustrates how players estimate their own status while playing a game, and how their partners perceive their status in two cases: when they are connected (i.e.~intra-clan) and when they are from another network component (i.e.~inter-clan).
Unsurprisingly, within clans these estimates are more closely aligned (i.e.~closer to the identity line)  than between individuals from different clans (0.005 vs.~0.017 average Euclidean distance to the diagonal).
\section{Discussion}\label{sec:discussion}
In this study, we examined the co-evolution of values, social hierarchy, social ties formation, and strategies in the UG. Our findings reveal that even within a relatively simple framework, diverse strategic behaviours emerge, ranging from highly greedy to exceptionally generous offers. These outcomes are primarily influenced by three key factors: (1) the age at which individuals become reproductively active, (2) the structure and formation of network connections, and (3) the social hierarchy, which is linked to observables of the game and influences the role allocation in the UG. In a modeled society where all players alternate roles equally each round (an assumption common in many models in the literature) and follow the same strategy, thresholds become less relevant, as everyone receives the same payoff. Introducing randomness in the role assignment, while keeping equal probability for playing each role, adds some variability to individual payoffs, however, the average payoff again remains unchanged. Models with network heterogeneity bring in inequality in average rewards, as higher-degree nodes are chosen more frequently by their neighbours, accumulating greater rewards, but again their average payoff \textit{per game} is the same for every player. The scenario where this no longer holds---and the main focus of our study--- is when role assignment is systematically imbalanced. In such cases, players assigned to the lower-earning role face a significant disadvantage, making thresholds critically important.

Another key characteristic that sets this model apart from most existing ones is its dynamically evolving population size, along with the requirement that a node must have an energetically sufficient and sufficiently old neighbouring node to reproduce. Research on growing networks has been conducted in the context of the Prisoner's Dilemma (see e.g.~\cite{poncela2009}) and also the UG (see e.g.~\cite{deng2012}). The studies demonstrate that this feature fosters cooperation in the former and altruism in the latter, compared to networks of fixed size. We see a similar effect, when we run a "typical" evolutionary-game theoretical model of UG on a structured network, where the topology is set as the evolved network in our simulation (green wiring). In that case we get significantly lower thresholds than the (almost) fair ones in our growing (and shrinking) network setup of Null value system. However, in the majority of classical models, network size remains fixed while structure can evolve through gradual relinking, e.g.~by relinking from low-offering players to a randomly selected player, thus promoting fairer thresholds (see e.g.~\cite{gao2011,miyaji2013}). In contrast, our model enforces a harsher penalty---when a node's energy reaches zero, it disconnects from all neighbours simultaneously and is removed, imposing severe consequences not only on poorly performing players but also on their entire neighbourhood (by loosing a partner to play with and reproduce). While this bears some resemblance to the social penalty in~\cite{sinatra2009}, where low-performing nodes and their neighbours reset strategies to random, our approach differs fundamentally as removed nodes do not re-enter the system. Additionally, new nodes are predominantly connected to their parents and the parents' neighbours (depending on the wiring scenario), who must be energetically rich and within a predefined age range. Deng et al.~\cite{deng2012}, for instance, show that the choice of nodes to which new nodes connect in their growing network significantly affects threshold levels---the stronger the tendency to connect to high-payoff nodes, the greater the resulting fairness.

The results differ markedly between the fully connected network and the structured network with initialised ring structure of degree 8. In the fully connected network with a minimal reproductive age 180, only populations with (almost) flat social hierarchies
survive, and when the age is set to zero, additionally also the Energy value systems survive (in five out of twenty simulations). This occurs because players in the fully connected network fail to learn to offer fair splits, which are crucial for preventing societal collapse in the face of unequal role allocation. However, in the structured network, players are able to learn this and commonly offer fair and generous splits. In contrast to the fully connected network, where a node loss (or creation) affects the entire population, in the structured network, the impact is localised---when a node dies, only its neighbours lose connections, reducing their chances of playing the UG and reproducing. Similarly, reproduction benefits a limited portion of the network, not all players. Successful reproduction requires strong neighbours, both energetically and in terms of age, creating clear incentives to offer splits that balance the fitness of the node and its neighbours.

For instance, in the Age value systems, younger players have a lower probability of being proposers compared to older players and typically have fewer connections. The number of connections peaks during middle age, as individuals create child nodes, and declines again in old age when connections begin to die out. Consequently, younger players, being the least frequent proposers, are at a disadvantage. If offer thresholds were too low, young players might die before reaching the reproductive age, leading to the collapse of the "community", creating an evolutionary incentive for increasing these thresholds above greedy values. If reproduction is allowed from age zero, being generous to younger players is less crucial, as long as child production outpaces the mortality rate, which is approximately ten times higher at reproductive age zero compared to 180--- similarly to average lifetime.
Another key property is the average clustering coefficient, which, as expected, correlates with the network's wiring--- greater connectivity between parent's neighbours and its child results in higher clustering, ranging from approximately 0.05 (blue wiring (0)) to 0.5 (green (6)). Previous work~\cite{kuperman2008} demonstrated that high clustering in the small-world networks promotes higher offer thresholds. We observe a similar relationship in our model. However, isolating the exact cause is challenging given the model's co-evolution of multiple factors. We also observe slight differences between clustering coefficients within the same wiring setup, but different value systems. 

When looking at the green wiring scenario, the child nodes form connections only with their parents' neighbours and the network gradually splits into isolated communities.  
For example, in the Energy value system, evolutionary thresholds shift toward configurations where the energetically wealthier player in a pair (proposer) allocates a larger portion of the reward to the energetically poorer partner. This promotes greater energetic equality in the network, reflected in lower Gini index values (see Figure~\ref{fig:distributions} and~\ref{fig:gini_idx}). However, as the setup transitions to the ones with more random connections, community structures begin to break down (Figure~\ref{fig:sw_network_structure}) and thresholds that once influenced only tightly knit communities now affect broader population, resulting in decreased generosity and increased inequality.

In situations where node death is possible and societal fitness relies on the survival and reproductive success of pairs of players, we suggest that certain fairness norms may be more group-beneficial than others, as they promote a more equitable redistribution of resources within unequal hierarchies (but see also~\cite{akdeniz2021}).

Ultimately, our model remains a strongly simplified representation of social behaviour. In reality, formation of social ties and life cycle are far more complex, sharing dynamics extend beyond the UG framework, and properties of the whole social community can play a significant role (see, e.g.~\cite{krakovska2025} for a model of UG with group interactions instead of pairwise ones). Nonetheless, our findings suggest that social hierarchy may play a crucial role in shaping the evolution of fairness, highlighting how values tied to reward distribution co-evolve within a dynamically growing and shrinking network.

\section{Conclusions}\label{sec:conclusions}
In this paper, we introduced a model of a dynamically growing/shrinking society that evolves over time---players age, reproduce, and die from old age or starvation. Rewards (i.e.~energy) are distributed via the UG, with role allocation determined by individual values and subjective status estimates derived from them. We explored various configurations of the value systems, reproductive conditions, and mechanisms for integrating newly added child nodes into the social network.

Our simulations revealed a wide spectrum of offering behaviors, ranging from greedy to fair, and even highly generous proposals similar to the ones found in small-scale societies~\cite{henrich2005}. These behavioral patterns were influenced by multiple factors, including the reproductive age window, the structure of connections formed by children, and the underlying social hierarchy, which played a key role in determining who assumed the proposer role in interactions.

In comparison to other models in the literature, our model does not require any form of reputation (e.g.~agents knowing what offers a player accepted/offered in the previous rounds~\cite{nowak2000}), spite or any preference between the players~\cite{debove2016}. The only ingredients that are in the model are realistic features of human society--- namely restricted age window of diploid reproduction, construction of social connections after birth, death of players (nodes) and values that guide the social hierarchy in the society. These are based on traditional observables such as age, energy (wealth), reproduction etc. The agents do not have memory and only know the basic observables of their partnering players. Their strategies remain fixed throughout their lifetime, without any adaptation. Although these choices simplify the model, they may also be seen as limitations that reduce its realism.

Our model does not explore the origins of values in the toy society, but rather their influence on sharing thresholds. However, in two systems---the All system and the A+E system---we allow the population to evolutionarily select the individual weights assigned to various values from a predefined set. In the All system, the Order value often functions as a proxy for a flat hierarchy configuration, as most players develop well-ordered thresholds. Nonetheless, the Order value system does not always emerge as the dominant evolutionary outcome. Different network structures give rise to diverse prevailing value systems---while many favour the Order value, which promotes flat hierarchy, others result in the dominance of Energy, Age, Good Acceptor, or a combination of multiple values. Moreover, whichever value system emerges, it tends to be widely adopted and internally coherent across the population.

Overall, our model highlights how simple, realistic societal features can drive the evolution of fairness in social interactions, illustrating the dynamic interplay between social hierarchy, resource distribution, and dynamically changing network structure (and size) in shaping sharing behaviour of resources.

Naturally, the reality of social and sharing dynamics is far more complex than what is captured by our model. Real-world societies involve multifaceted value systems, overlapping hierarchies, and context-dependent interactions that evolve over time. Our approach is a simplification, intended to isolate key mechanisms and explore their effects in a controlled setting. 

\section*{Acknowledgments}
This project has received funding from the European Union’s Horizon 2020 research and innovation programme under grant agreement No.~955708. HK also acknowledges support from the Slovak Grant Agency for Science (VEGA 2/0023/22). 
\bibliography{references}
\end{document}